\documentclass[twocolumn]{aastex701}
\usepackage{graphicx} 

\usepackage{rotating}

\usepackage{xcolor}

\begin{document}

\title{Discovery of Very Minor Merger-Induced Star-forming Clumps at $z=3.33$ with JWST}

\correspondingauthor{Alexander de la Vega}

\author[0000-0002-6219-5558, gname=Alexander, sname='de la Vega']{Alexander de la Vega}
\affiliation{Department of Physics and Astronomy, University of California, 900 University Ave, Riverside, CA 92521, USA}
\email{alexandd@ucr.edu}

\author[0000-0001-5846-4404]{Bahram Mobasher}
\affiliation{Department of Physics and Astronomy, University of California, 900 University Ave, Riverside, CA 92521, USA}
\email{mobasher@ucr.edu}

\defcitealias{delaVega26}{dlV26}



\begin{abstract}

\noindent High-redshift galaxies often contain ``clumps:'' compact, $\sim$kpc-scale regions of intense star formation. These substructures make up significant fractions of the masses and star-formation rates of their host galaxies and are expected to affect the morphologies of their hosts, such as by building up bulges. Studying clumps can therefore inform theoretical models of galaxy evolution. Two main mechanisms have been proposed for their formation: gas fragmentation from disk instabilities and compression of gas during gravitational interactions with other galaxies. In the latter, only mergers with mass ratios greater than 1:10 have been considered. Here, we report James Webb Space Telescope spectroscopy and imaging of JADES 1034862, a galaxy at $z_{\textrm{spec}} = 3.33$, which hosts a clump that formed during an interaction with a companion with a stellar mass that is $\approx28$ times lower than the host. From fits to the spectral energy distributions of the clump and the companion, it is found that both of their star-formation histories rose sharply in the most recent 100--300 Myr, during which the clump formed $\sim10^{7.2} \ M_{\odot}$ in stars and currently makes up $\approx10\%$ of the total star-formation rate of its host. Given the higher frequency of mergers with low mass ratios ($<1:10$) to those with higher ratios, we infer that a substantial fraction of high-redshift clumps could have formed through weak gravitational interactions.
\end{abstract}


\section{Introduction}

Galaxies at intermediate and high redshift $(z\gtrsim1)$ often exhibit ``clumps:'' compact ($\lesssim1$ kpc) regions of intense star-formation, which make up significant fractions of the total star-formation rates (SFRs) and stellar masses in these systems \citep{Giavalisco96, Elmegreen07, ForsterSchreiber11, Genzel11, Guo12, Guo15, Guo18, Shibuya16, Soto17, Sattari23, Kalita25, delaVega26}. Clumps are believed to form mainly through processes internal to their hosts (in situ) or externally (ex situ). In-situ mechanisms include violent disk instabilities \citep{Noguchi99, Immeli04_hst, Bournaud07, Agertz09, Dekel09_theory, Ceverino10} and turbulent star-formation \citep{Sun26}, while ex-situ mechanisms include compression of gas during major mergers \citep{DiMatteo08, Renaud15, Li17, Nakazato24} and accretion of satellite galaxies by their hosts \citep{Mandelker14, Zanella19}. Regardless of their origin, clumps are expected to drive strong morphological changes in their host galaxies, such as migrating to the centers to form bulges \citep{Noguchi99, Bournaud07, Dekel09_theory, Bournaud14, Mandelker14} and/or dissipating rapidly and producing strong gaseous outflows \citep{Murray10, Genel12, Hopkins12, Mandelker17}. 

Both theoretical and observational studies have so far pointed to major or minor mergers, with mass ratios $\geq$1:10 \citep[e.g.,][]{Lotz10}, as the main processes responsible for the formation of merger-induced clumps. Such mergers have been identified as significant clump formation processes at low redshift \citep[$z\lesssim1$,][]{Puech10, Calabro19}, in massive and/or ultraviolet (UV)-bright galaxies over $2<z<7$ \citep{Ribeiro17, Bik24, Tanaka24, Mawatari26}, and at low stellar masses \citep[$9<\log\left(M_{\star}/M_{\odot}\right) \lesssim10$,][]{Guo15, delaVega26}. However, mergers at lower mass ratios, so-called ``very minor mergers,'' \citep[see ][]{Rodriguez-Gomez16} occur more frequently \citep{LaceyCole93, Maller06, Bottrell24}.  Whether clumps can form from such mergers has not been thoroughly explored. 

In this work, we show that clumps may also form via minor gravitational interactions at a significantly lower mass ratio ($\approx$1:28) than heretofore examined. We present a system consisting of a clumpy galaxy, JADES 1034862, and its companion, JADES 1034865, both at redshift $z\approx3.33$. This system provides an example showing how very minor mergers could induce bursts of star-formation activity, leading to the formation of clumps. This paper is organized as follows. In Section \ref{sec:data_sample}, we describe the data and selection of JADES 1034862. In Section \ref{sec:spectra}, the analysis performed on JWST spectra is described, including the measurement of spectroscopic redshifts and evidence that JADES 1034862 and its companion are interacting. In Section \ref{sec:detect_clumps}, we describe how clumps are detected and photometry is performed. The star-formation histories of the clumps, their host, and its companion are examined in Section \ref{sec:sfhs}. We summarize and discuss our findings in Section \ref{sec:discussion}. 

We adopt cosmological parameters measured by \citet{Planck16}, with $H_0 = 67.7~\rm{km}~\rm{s}^{-1}~\rm{kpc}^{-1}, \Omega_{\rm{m}} = 0.309$, and $\Omega_{\Lambda} = 0.691$. Magnitudes are expressed in the AB photometric system \citep{Oke83}. 

\begin{figure*}[t!]
    \centering
    \includegraphics[width=\linewidth]{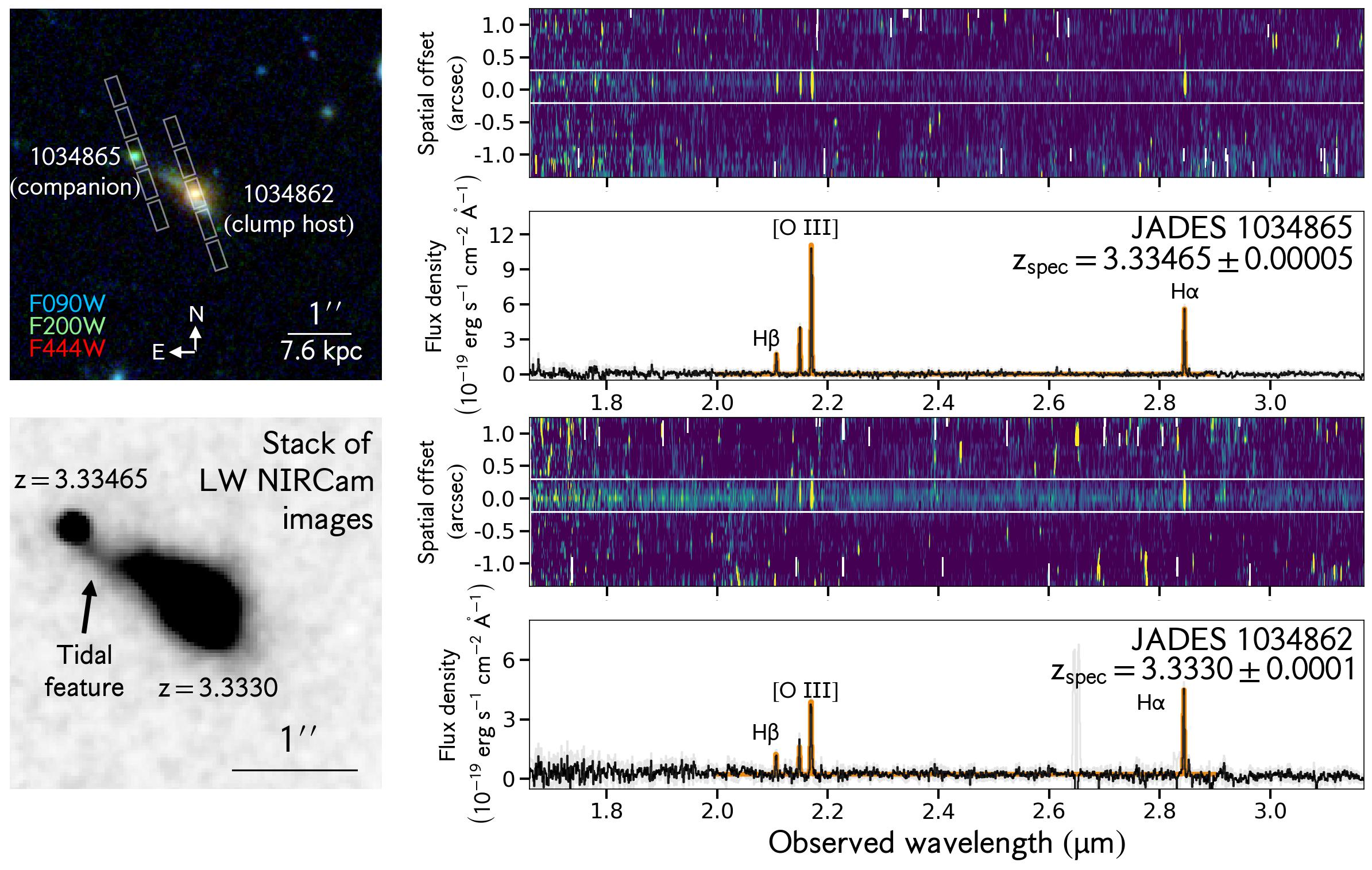}  
    \caption{NIRCam RGB image (top left panel), a zoomed-in stack of long-wavelength (2.77 -- 4.44 $\mu$m) NIRCam images (bottom left) and NIRSpec spectra (right) of the target and companion. The target (JADES ID 1034862) is located in the center of the RGB image and its companion (JADES 1034865) is to the upper left, to the northeast, $\sim8$ kpc away. A tidal feature connects them, as shown in the stacked image. To determine spectroscopic redshifts, medium-resolution spectra from JADES are used. The locations of the Micro-Shutter Assembly shutters are shown for each source as white rectangles. The two-dimensional spectra are shown in the top panels on the right. One-dimensional spectra (bottom panels) are extracted in the regions indicated with the white lines in the top panels. Spectral flux densities are shown in black and errors in faint gray lines. Redshifts are measured by simultaneously fitting a flat continuum and Gaussian profiles to H$\beta$, [O {\sc iii}], and H$\alpha$, shown as thick orange lines. The redshifts are also given in the stacked image next to each galaxy.}
    \label{fig:rgb_img_spectra}
\end{figure*}

\section{Data and Selection of Targets}
\label{sec:data_sample}

The targets were selected from NIRCam and NIRSpec observations of the Great Observatories Origins Deep Survey \citep[GOODS;][]{Giavalisco04} North field (hereafter, GOODS-N), taken as part of Data Release 3 of the JWST Advanced Deep Extragalactic Survey \citep[JADES;][]{Rieke23, Bunker24, DEugenio25, Eisenstein25, Eisenstein26}. The GOODS-N field was observed in 11 NIRCam bandpasses: F090W, F115W, F150W, F182M, F200W, F210M, F277W, F335M, F356W, F410M, and F444W. In each bandpass, the 5$\sigma$ point-source limiting magnitude measured in 0.\arcsec3 circular apertures is $\sim29.5$ mag. We use Kron photometry to fit the spectral energy distributions (SEDs) of the targets, with a Kron parameter of $k=2.5$. Spectroscopic observations were taken with the NIRSpec Micro-shutter Assembly (MSA) and the targets lie in the `mediumhst' tier (Program ID 1181). Spectra taken with the G235M grating are used, with resolving power $R\sim1000$. A three-point nodding pattern was used, with an an exposure time of 3107 s. We refer the reader to \citet{Rieke23} and \citet{Eisenstein26} for more details on the NIRCam imaging and photometry and to \citet{Bunker24} and \citet{DEugenio25} for more details on the NIRSpec observations and data reduction. 

The targets were selected from a sample of $\sim9000$ star-forming, clumpy galaxies over $2<z<12$ identified in \citet[][hereafter, \citetalias{delaVega26}]{delaVega26}. The selection of a parent sample of galaxies and a subsample of clumpy galaxies was given in that work. To these cuts, we further select clumpy host galaxies that have neighbors with projected separations $<30$ kpc. Both the hosts and neighbors must have spectroscopic redshifts from JADES and line-of-sight velocity differences $<1000$ km s$^{-1}$. This search resulted in 15 galaxy pairs. Of these, JADES 1034862 (R.A. = 189.11680, Decl. = 62.29810, NIRSpec ID 1233) and 1034865 (1240) make up the pair with the highest stellar mass of the primary (a ratio of $\approx$1:28) and smallest separation ($<10$ kpc). The galaxies and their spectra are shown in Figure \ref{fig:rgb_img_spectra}, displaying evidence of undergoing a merger. 

\section{Spectroscopic Analysis}
\label{sec:spectra}

To demonstrate that the clump host and its companion are interacting, we measure their spectroscopic redshifts and calculate a velocity difference. This is described below.

\subsection{Extraction of One-Dimensional Spectra}

One-dimensional (1D) spectra are used to determine redshifts. While 1D and 2D spectra were included in Data Release 3 of JADES, we opt to extract 1D spectra for our targets. This is because the released 1D spectrum of JADES 1034862 contains a bright artifact at $\approx2.65~\mu$m that could be mistaken for an emission line. It lies about one shutter height ($\approx$0.\arcsec4) above the trace. It ended up in the 1D spectrum because the released 1D spectra were created via a weighted combination of individual 1D spectra extracted from each of the three nod positions \citep[see Sec. 4 in][]{Bunker24}. 

We extract the 1D spectrum for each galaxy as follows. The 2D spectrum is visually inspected for the trace. An extraction aperture of width equal to 5 pixels (0.\arcsec5) is defined and centered on it. For each wavelength channel, the extracted flux density is taken to be the sum over all pixels and the extracted error is the sum of the pixel errors in quadrature. This assumes that errors are uncorrelated. It is corrected by rescaling the error spectrum to the noise fluctuations in the science spectrum, following previous studies \citep[e.g.,][]{Ubler2023, Lamperti24, Fujimoto25}. This is done by calculating the standard deviation as a function of wavelength from the science spectrum using a window of 11 pixels. Regions with emission lines are masked. The error spectrum is accordingly rescaled and added to the original error spectrum. This results in a final error spectrum that is larger than the original by a factor of $2.2-2.6$, on average.

\begin{figure*}[t!]
    \centering
    \includegraphics[width=\linewidth]{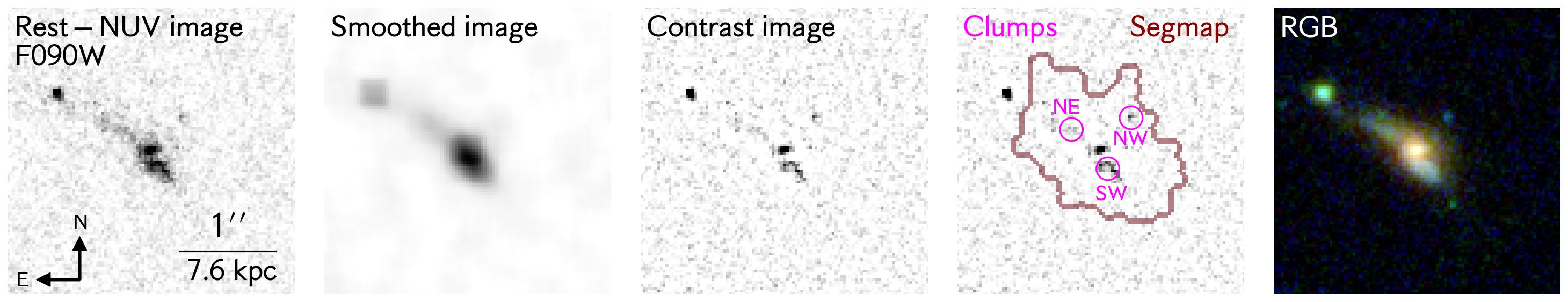}  
    \caption{The procedure used to find clumps in the target. Clumps are searched for in an image spanning the rest-NUV (leftmost column). It is smoothed (second from left) and subtracted from the original to produce a contrast image (middle). Source detection is performed on the contrast image and high-S/N candidate sources are taken to be clumps (magenta circles in second from right panel). Clumps must lie off-center and within the segmentation map, shown as a red line. Clumps are blue regions within their host, as seen in the RGB image (right).}
    \label{fig:clump_detection}
\end{figure*}

\subsection{Fitting Spectroscopic Redshifts}
For each galaxy, its redshift is measured by fitting the H$\beta$, [O {\sc iii}] $\lambda\lambda4960, 5008$, and H$\alpha$ emission lines simultaneously to its 1D spectrum. This is done by assuming a Gaussian profile for each line and a continuum that is constant with wavelength. This model is fit using {\sc pymultinest} \citep{Buchner14}, the python implementation of the Bayesian Markov Chain Monte Carlo algorithm {\sc multinest} \citep{Feroz09}. The redshift varies freely within the range [3.3, 3.4]. All other parameters (amplitudes and widths of Gaussians and value of the continuum) are left to vary freely, with two exceptions. The ratio of the amplitudes of [O {\sc iii}] $\lambda5008$ to [O {\sc iii}] $\lambda4960$ is fixed to 2.98 \citep{Storey00} and the two lines share the same width.  A Gaussian likelihood is assumed. The model is fit to the portion of the spectrum spanning 2.0 to 2.9 $\mu$m in the observed frame. 

Fits are shown in Figure \ref{fig:rgb_img_spectra}. Given the high quality of the spectra, the redshifts are measured precisely: the redshift of the clump host is $z_{\rm{host}}=3.3330\pm0.0001$ and that of the companion is $z_{\rm{companion}}=3.33465\pm0.00005$. 

\subsection{Evidence that the Clump Host and the Companion Satellite Are Interacting}

Three pieces of evidence demonstrate that the two galaxies are interacting: a small line-of-sight velocity difference, a small projected separation, and a faint tidal feature linking them. The former is given by $\Delta v = c\times\frac{\Delta z}{(1+z_{\rm{host}})},$ where $\Delta z = |z_{\rm{host}} - z_{\rm{companion}}|$. We find $\Delta v = 114\pm8$ km s$^{-1}$, where the uncertainty is derived through propagation of error. The projected separation, $\Delta r$, is 1.1\arcsec, which is 8.7 kpc at $z_{\rm{host}}$. These values are much smaller than limits that have been typically used to select gravitationally bound galaxy pairs \citep[$\Delta v \leq 300-1000$ km s$^{-1}$ and $\Delta r \leq 150$ kpc, e.g.,][]{Patton00, Patton13, Scudder12, Shah20}. Last, there is a faint tidal feature linking the two galaxies. This is seen in Figure \ref{fig:rgb_img_spectra} in a stacked image of the long-wavelength NIRCam bandpasses (2.77 -- 4.44 $\mu$m). 

\section{Identification and Photometry of Clumps}
\label{sec:detect_clumps}

Clumps in this system were identified following the procedure described in Sec. 4 in \citetalias{delaVega26}. In that work, clumps were selected to be complete in UV brightness. Below, we describe how they were detected. 

\begin{figure}
    \centering
    \includegraphics[width=0.9\linewidth]{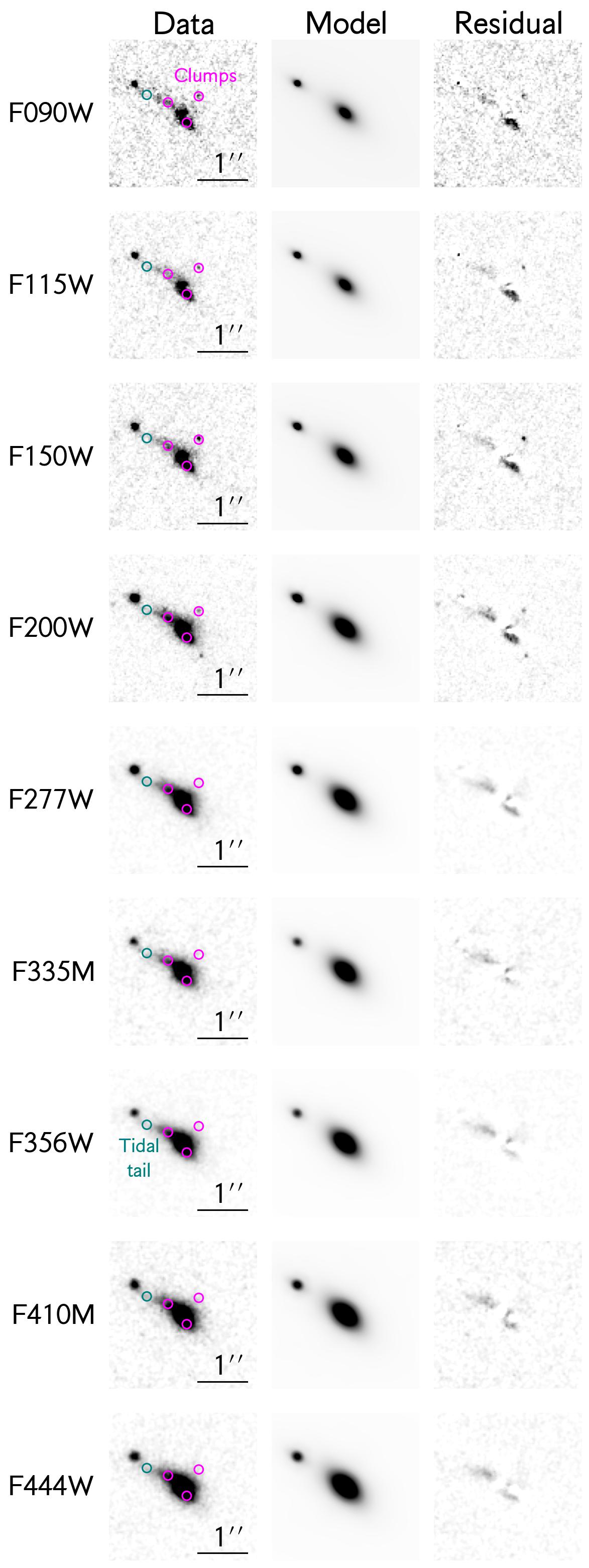}  
    \caption{Morphological fits to the clump host and companion in nine NIRCam bandpasses. Each galaxy is fit with a one-component S\'{e}rsic model in all nine bandpasses simultaneously using the {\sc galfitm} tool. The original data, models, and residual images are shown in the left, middle, and right columns, respectively. The clumps identified in Figure \ref{fig:clump_detection} and a faint tidal feature are masked during the fit. The former are indicated with magenta circles and the latter with a teal circle. The tidal feature is best seen in F277W and redder filters. The fit to the clump host is used to model the diffuse background when performing photometry on the clumps (Section \ref{sec:detect_clumps}).}
    \label{fig:galfitm_fit}
\end{figure}

Clumps are identified in the rest-NUV, which corresponds to F090W at the redshift of the clump host. The image of the galaxy in this bandpass is smoothed with a boxcar filter of size equal to 10 pixels. The smoothed image is subtracted from the original to create a contrast image. Source detection is performed on the contrast image to find candidate clumps. Next, deblending is performed. Candidates must have signal-to-noise ratios (S/Ns) $\geq 3$ and span at least 5 contiguous pixels, all of which must be at least two standard deviations brighter than the background of the contrast image. Finally, to be selected as clumps, candidates must lie at least 0.\arcsec1 from the center of the host and within the segmentation map provided by the JADES team. 

In order to select a complete sample of clumps, their brightnesses were measured in the rest-NUV. This is done with aperture photometry. Circular apertures of radius 4 pixels are centered on the clumps and the local background is measured in circular annuli with radii $6-10$ pixels. The mean fluxes within the annuli are calculated and multiplied by the area of the central apertures. The expected background fluxes are subtracted from those within the central apertures. An aperture correction of 1.511 is applied to account for the extended morphologies of the clumps. A minimum fractional luminosity of 3.5\% of the host's light in F090W is used to select a complete sample of clumps, given the host's stellar mass of $\log\left(M_{\star}/M_{\odot}\right) \approx 10$ (Section \ref{sec:sfhs}).

Three clumps are identified and shown in Figure \ref{fig:clump_detection}. They lie to the northeast, northwest, and southwest of the center of the clump host and will be referred to as the NE, NW, and SW clumps hereafter, respectively. To measure clump SEDs, we perform another round of aperture photometry on point spread function (PSF)-matched images with the background light of the host removed. Empirical PSFs from \citetalias{delaVega26} are used. 

 The light profile of the host galaxy is modeled with 2D S\'{e}rsic functions \citep{Sersic68}. This is done to subtract the background light of the host and more accurately measure the SEDs of clumps. We use the {\sc galfitm} tool \citep{Barden12, Haussler13, Haussler22, Vika13}, which is based on {\sc galfit} \citep{Peng02, Peng10}, to fit S\'{e}rsic functions that vary with wavelength to all nine NIRCam bandpasses listed in section \ref{sec:data_sample}. We adopt the setup developed in \citet{Ward24} and assume third-order Chebyshev polynomials of the first kind \citep{AbramowitzStegun65} to model the effective radius and S\'{e}rsic index as a function of wavelength. Following \citet{Nedkova21} and \citet{Ward24}, we fit the axis ratio, position angle, and central coordinates each as constant values with wavelength. The magnitudes of the models are left to vary freely in each bandpass. Each of the three clumps is masked with a circle of radius 3 pixels. The center of the tidal feature linking the two galaxies is also masked with a circle of the same size. The best-fit model and residual images are shown in Figure \ref{fig:galfitm_fit}. The fit adequately describes the morphology of the host and resulted in a reduced $\chi^2$ of 0.8 across all bandpasses. Over the wavelength range considered here, the effective radius of the clump host is $R_{\rm{eff}} = 0.27$\arcsec $-0.35$\arcsec and its S\'{e}rsic index is $n=1.5-2$. 

Photometry on the clumps is performed as follows. First, all NIRCam images are convolved to the resolution of F444W using PSF-matching kernels created with {\tt pypher} \citep{Boucaud16}. Next, the S\'{e}rsic models described above are convolved to match the F444W PSF and subtracted. Circular apertures of radius equal to 4 pixels are placed at the positions of the clumps. Their fluxes in all bandpasses are measured from the difference images between the observed and best-fit Sersic models. Background correction is done using circular annuli as described earlier in this section. This corrects for residual host light not accounted for by the model and is generally $\lesssim10\%$ of the clump light. Last, an aperture correction of 2.046 is applied to the background-subtracted fluxes of clumps. This is calculated by assuming that the light profiles of clumps here largely follow the typical S\'{e}rsic profile ($R_{\rm{eff}} = $0.\arcsec046, $n=1.04$) of clumps found in \citetalias{delaVega26}. The profile is convolved with the F444W PSF in GOODS-N and the correction is measured for a circular aperture of radius equal to 4 pixels. Errors on the clump fluxes are taken to be sums in quadrature of the pixel errors within the circular apertures multiplied by the aperture correction. SEDs of the clumps and their errors are shown in Figure \ref{fig:sed_fits}. 

\section{Star-Formation Histories of the Clumps and the Merging System}
\label{sec:sfhs}

Star formation histories (SFHs) are obtained by fitting the SEDs of the clumps, their host, and the companion with the Bayesian tool {\sc bagpipes} \citep{Carnall18}. We adopt the priors and their ranges provided in section 2.3 of \citetalias{delaVega26}, with some exceptions as noted. Briefly, we assume physically motivated priors on the metallicity and the attenuation $A_V$ as noted by \citet{Leja19_3dhst} and \citet{delavega25}. The flexible attenuation curve model by \citet{Salim18} is assumed. Differential attenuation between young ($\leq10$ Myr) and older stars is applied, following \citet{CF00}. 

\begin{figure*}[t!]
    \centering
    \includegraphics[width=\linewidth]{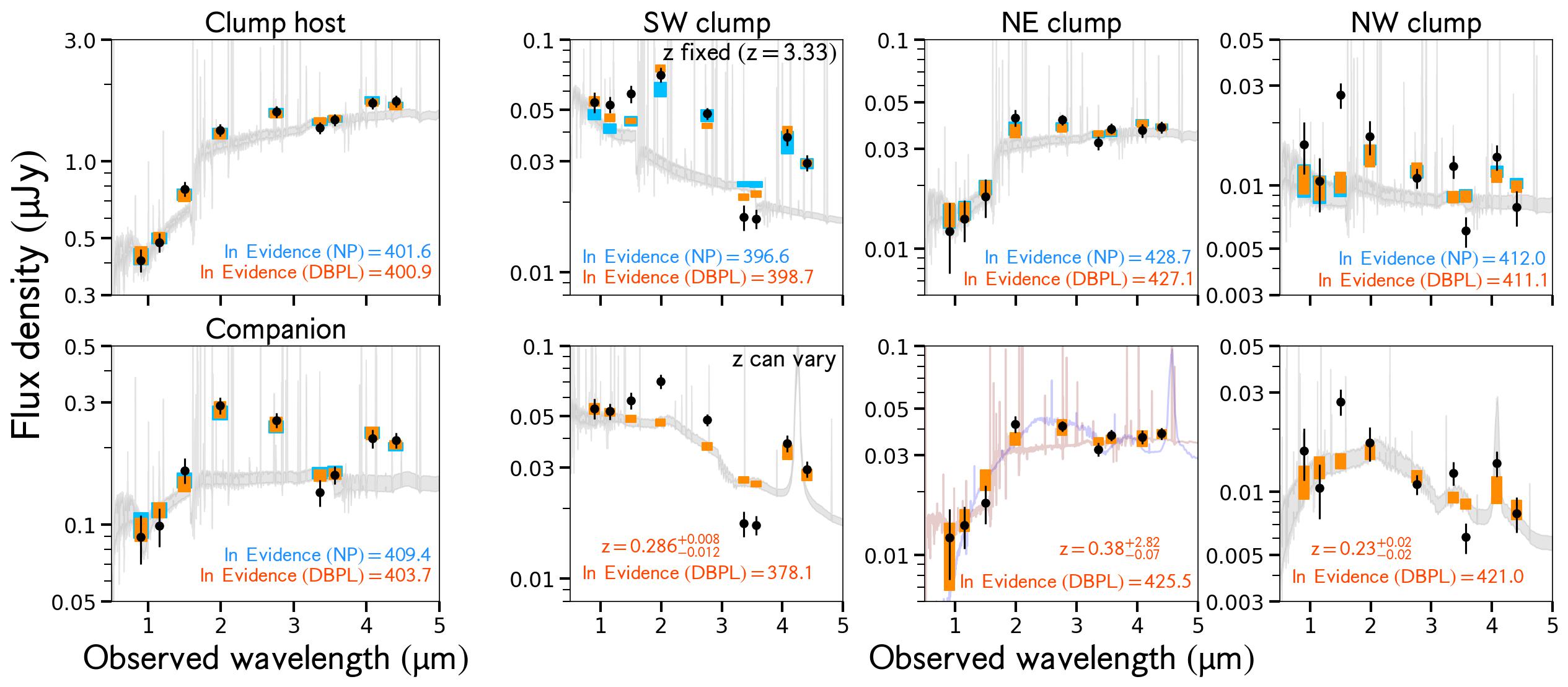}  
    \caption{Spectral energy distribution fits to the target and its companion (left column) and the clumps (middle and right columns). Observed NIRCam photometry is shown as black symbols with error bars. Each object is fit with non-parametric (shown in blue) and double power-law SFHs (shown in orange). Model photometry from the fits is shown as small squares with their heights spanning the 16th to 84th percentiles from the fit posterior distributions. For the galaxies on the left, fits are performed with the redshifts fixed to the spectroscopic values. For the clumps, two sets of fits are performed: one with the redshift fixed to $z=3.33$ (top row) and another with redshift left to vary (bottom). In the lower half of each panel, the value of the Bayesian evidence under each model is given. When varying the redshift, two redshift solutions are found for the NE clump, one at $z\approx0.38$ (blue line) and another at $z\approx3.2$ (red line). }
    \label{fig:sed_fits}
\end{figure*}

In this work, we consider two types of SFHs: non-parametric and parametric. While non-parametric SFHs more accurately recover the presence of older stars in the SEDs, which affects the resulting inferred stellar masses and ages \citep{Leja19_np_sfh, Lower20, Jain24}, they generally require more free parameters to fit. Therefore, the parametric SFH is included for comparison.  In the non-parametric SFH, the SFR varies as a piecewise function across several bins in lookback time. In each bin, the SFR is a constant. We adopt the ``continuity'' prior from \citep{Leja19_np_sfh} and use eight bins in lookback time: $0<t<10$, $10<t<30$, $30<t<50$, $50<t<100$, $100<t<300$, $300<t<500$, $500<t<1000$, and $1000<t<1500$ Myr. The parametric SFH is given by a double power-law \citep{Carnall18}, as was used in \citetalias{delaVega26}. 

The latest version of the stellar population synthesis models by \citet{BC03} is used. Nebular emission models were obtained using {\sc cloudy} \citep{Ferland17}. The strength of the nebular emission is given by the ionization parameter, which varies freely. The initial mass function from \citet{Chabrier03} is adopted. 

The impacts of choosing these model assumptions are examined as follows. The SEDs of the clump host and its companion are fit with the non-parametric and double power-law SFHs. The redshifts are fixed to their spectroscopic values. The SEDs of the three clumps are each fit with both types of SFH. in one set, their redshifts are fixed to $z=3.33$. In another, their redshifts are allowed to vary over the range [0, 10] and are fit with only the double power-law SFH. The latter is done to determine whether all clumps lie within the host or are interlopers at other redshifts that lie at close projected separation. 

The fits are shown in Figure \ref{fig:sed_fits}. For each fit, the natural log of the Bayesian evidence is recorded. This is the product of the likelihood and prior that is integrated over the whole parameter space. Two models (e.g., two types of SFH) may be compared by calculating the ratio of their Bayesian evidences \citep[e.g.,][]{HanHan12, HanHan14, HanHan19, Chevallard16, Salmon16, delavega25}. To interpret the strength of the preference of one model to another, the criteria from \citet{KassRaftery95} are adopted, in which preferences are categorized as ``weak,'' ``positive,'' ``strong,'' or ``very strong.''

The non-parametric SFHs are generally preferred over the double power-law ones. The former is very strongly preferred over the latter for the companion, and is weakly preferred for the host and its clumps. For the SW and NE clumps, fits with $z=3.33$ are very strongly preferred to those in which the redshift varies. Under the non-parametric SFHs, the stellar masses of the clump host, companion, and SW and NE clumps are $\log\left(M_{\star}/M_{\odot}\right) = 9.99^{+0.06}_{-0.09}$, $8.55^{+0.15}_{-0.15}$, $7.22^{+0.03}_{-0.03}$, and $8.34^{+0.06}_{-0.05}$, respectively. The values here are taken to be the medians from the posteriors and upper and lower values are the differences between the medians and the 84th and 16th percentiles. The NW clump is much better fit with a freely varying redshift and is taken to be a low-redshift ($z\approx0.2$) interloper. 

\begin{figure*}[t!]
    \centering
    \includegraphics[width=\linewidth]{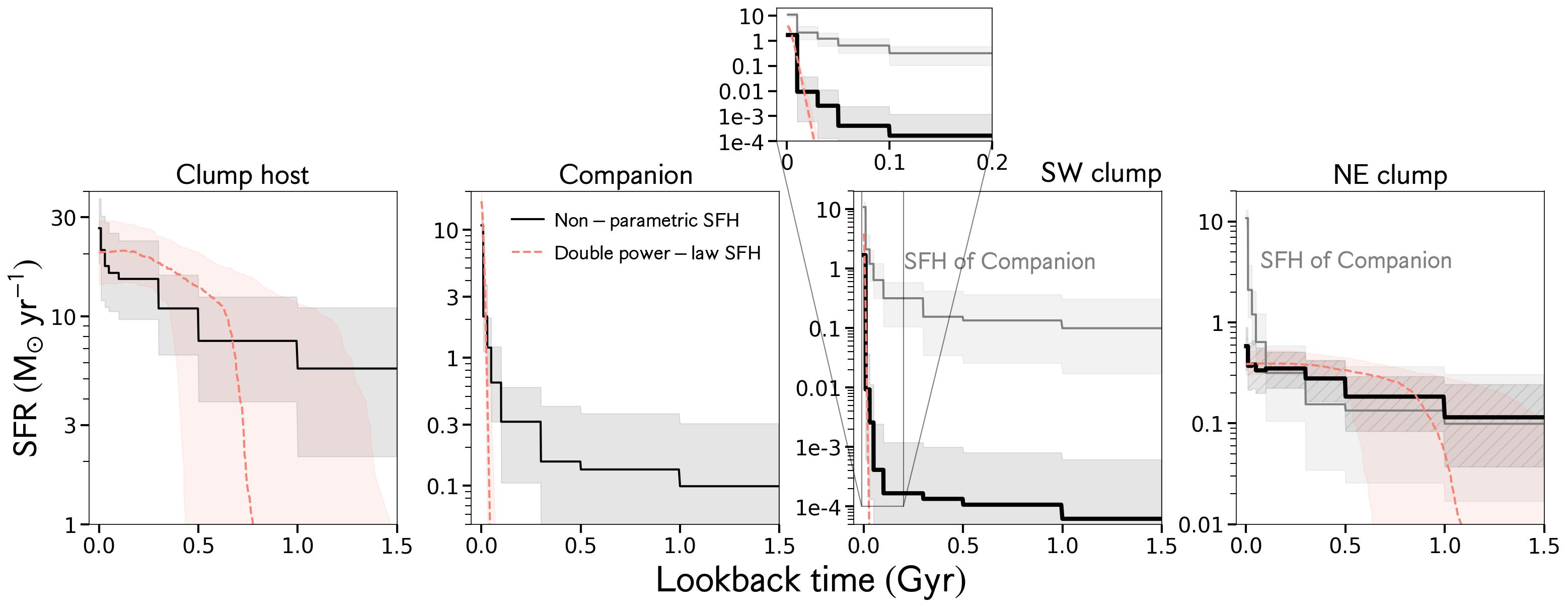}
    \caption{Star-formation histories (SFHs) of the clump host, companion, and SW and NE clumps, shown from left to right, respectively. In each panel, non-parametric SFHs are shown as solid black lines and double power-law ones as dashed light-red lines. Uncertainties on the SFHs are plotted as shaded regions with corresponding colors. The SFH of the companion is shown in light gray alongside those of the clumps (bold solid lines). The inset panel above the middle right one zooms in on the SFHs of the companion and SW clump.}
    \label{fig:sfhs_fit}
\end{figure*}

We now examine the inferred SFHs, shown in Figure \ref{fig:sfhs_fit}. As the lookback time decreases, the clump host's SFH rises gradually from $\sim3 M_{\odot}$ yr$^{-1}$ to $\sim20-30 M_{\odot}$ yr$^{-1}$ over $\sim1$ Gyr. The companion's SFH increases sharply from $\sim0.1$ to $
\sim10 M_{\odot}$ yr$^{-1}$ within the most recent $100-300$ Myr. The SFH of the SW clump also rose rapidly over the same timescale, from $\sim10^{-4}$ to $\sim2-3 M_{\odot}$ yr$^{-1}$. In contrast, the SFH of the NE clump increased much less rapidly over the same range within lookback time, from $\sim0.3$ to $\sim0.6 M_{\odot}$ yr$^{-1}$.

\section{Discussion}
\label{sec:discussion}

We discovered a merging system containing UV-bright clumps associated with a star-forming galaxy at $z=3.33$ using NIRCam imaging and NIRSpec spectra from the JADES survey. The clump host is $\approx28$ times more massive than its companion, meaning their interaction is a very minor merger. Nevertheless, it creates a tidal feature. The SFHs of the clump host, its companion, and clumps are inferred from fits to their NIRCam SEDs. The companion and the SW clump have SFHs that sharply rise (by $>2$ dex) in the last $100-300$ Myr, whereas those of the host and the NE clump change little (by $\lesssim0.3$ dex) over the same time range. 

The above suggests that the interaction between the companion and the clump host triggered the formation of the SW clump. Three pieces of evidence support this: the similar shape of the SFHs of the companion and the SW clump; the bluer color of the SW clump compared with the NE clump, suggesting a greater abundance of young stars in the former; and the lower stellar mass of the SW clump, suggesting a more recent time of formation. While minor mergers (with mass ratios between 1:4 and 1:10) have been posited as clump formation mechanisms \citep{Guo15, Shibuya16}, in this work, we show that very minor mergers can also form clumps. 

While we do not have enough data to constrain the history of the interaction, we can estimate what happened based on the observed morphology and inferred SFHs. Because the companion is to the NE of the host and the tidal tail connecting them extends along the same direction, we surmise that the companion may have traveled from the SW to the NE in front of or behind the host. As the companion can still be clearly distinguished from the host, we expect that the first pericenter passage of the interaction had just occurred. We estimate that the interaction began $\sim100-300$ Myr prior to $z=3.33$, since the SFHs of the companion and SW clump are relatively flat until those lookback times. The region of the host containing the SW clump may have been close to the companion early in the interaction, whereas the NE clump was farther away. Over time, as the companion moved to the NE, tidal forces during the interaction would have compressed gas in the SW clump, resulting in a rising SFH. The host could have rotated throughout the interaction, leading to the observed positions of the clumps, with the SW one on the farther side of the host from the companion.

The scenario we propose is consistent with simulations of starbursts during very minor mergers. We find that the SFHs of the companion and the SW clump are still rising $\sim300$ Myr after the first pericenter passage. Further, at the time of observation, $\approx10\%$ of the host's total SFR comes from the SW clump, which amounts to a SFR enhancement of $\sim0.05$ dex from just this clump. Using the cosmological TNG50 simulation \citep{Nelson19, Pillepich19}, \citet{Bottrell24} found that individual ``mini-mergers'' (mass ratios between 1:10 and 1:100) over $0.1<z<0.7$ can drive SFR enhancements of $\lesssim0.1$ dex in the 2 Gyr before coalescence. We note that SFR enhancements of $\lesssim0.3$ dex have also been found in simulated \citep{Moreno15, Fensch17} and observed \citep{Wild14, Weaver18} major mergers within $100-300$ Myr after the first pericenter passage. 

We note that the stellar mass of the companion is similar to that of the NE clump. Since clumps of similar mass are known to exist at $z\sim3$ \citep[e.g.,][]{Zhu26}, there is a possibility that the companion could instead have been a clump that was ejected from the host. Integral field spectroscopy could resolve this ambiguity. 

We conclude that very minor mergers are a plausible formation mechanism of massive $\left( \log\left(M_{\star}/M_{\odot}\right) \gtrsim 7 \right)$, star-forming clumps in high-redshift galaxies. Given the higher frequency of very minor mergers relative to their more massive counterparts \citep{Bottrell24}, it is possible that a non-negligible fraction of clumps formed this way. To determine this, large samples of spectroscopic redshifts must be obtained over a wide range in redshift and stellar mass, coupled with deep imaging needed to detect faint clumps. This is now possible with deep JWST surveys, such as JADES, which will form the basis of more comprehensive and systematic searches to come. 

\begin{acknowledgments}
    A.d.l.V thanks Niloofar Sharei for preparing beautiful cartoons of the interaction between the clump host and companion for use in talks and John Weaver for stimulating discussions on merging galaxies and inferring their SFHs from SED fits. 

    We thank the JADES team for designing and preparing their observations and releasing a rich public dataset. We also thank the STScI staff for enabling this science.

    This research has made use of the SVO Filter Profile Service ``Carlos Rodrigo" \citep{Rodrigo12, Rodrigo20}, funded by MCIN/AEI/10.13039/501100011033/ through grant PID2023-146210NB-I00.

    All of the JWST data used in this paper can be found in {\it MAST} \citep{Rieke23_doi}. 
\end{acknowledgments}

\software{astropy \citep{2013A&A...558A..33A,2018AJ....156..123A,Astropy22}, 
matplotlib \citep{Hunter:2007},
numpy \citep{harris2020array},
photutils \citep{Bradley20, Bradley23},
scipy \citep{2020SciPy-NMeth},
TOPCAT \citep{Taylor05}
          }

\bibliography{references}{}
\bibliographystyle{aasjournalv7}

\end{document}